\documentclass[superscriptaddress,floatfix,prl,twocolumn,amsmath,amssymb]{revtex4-1}
\usepackage{graphicx}
\usepackage{xcolor}
\usepackage{physics}
\usepackage[]{changes}
\usepackage[caption=false]{subfig}
\usepackage{multirow}

\graphicspath{{./Figures/}}

\begin{document}
\newcommand{\nvplus}{NV$^{+}$}
\newcommand{\nvminus}{NV$^{-}$}
\newcommand{\nvzero}{NV$^{0}$}

\title{Bright ab-initio photoluminescence of \nvplus~in diamond}
\author {Akib Karim}
    \email[Electronic address: ]{akib.karim@rmit.edu.au}
    \affiliation{Quantum Photonics Laboratory and Centre for Quantum Computation and Communication Technology, School of Engineering, RMIT University, Melbourne, Victoria 3000, Australia}
\author{Igor Lyskov} 
\author{Salvy P. Russo}
\affiliation{ARC Centre of Excellence in Exciton Science, School of Science, RMIT University, Melbourne, VIC
3001 Australia}
\affiliation{Chemical and Quantum Physics, School of Science, RMIT University, Melbourne VIC 3001, Australia}
\author{Alberto Peruzzo}
\email[Electronic address: ]{alberto.peruzzo@rmit.edu.au}
\affiliation{Quantum Photonics Laboratory and Centre for Quantum Computation and Communication Technology, School of Engineering, RMIT University, Melbourne, Victoria 3000, Australia}

\begin{abstract}

The positively charged nitrogen vacancy (\nvplus) centre in diamond has been traditionally treated as a dark state due to the experimental lack of an optical signature. Recent computational studies have shown that it is possible for the \nvplus\ defect to have an excited state transition equivalent to that of the negatively charged (\nvminus) centre, but no PL predictions have been reported so far. We report the first \textit{ab-initio} calculation showing that the \nvplus\ center presents quantum emission, with zero phonon line at 765\,nm and a non-zero transition dipole moment, approximately one quarter of the transition dipole moment of \nvminus. We calculate the energy levels of the multielectron states under time-dependent density functional theory (singlet and triplet E states), and, using our recently developed frequency cutoff method, we predict the full PL spectrum. Our results suggest that this state cannot be considered intrinsically `dark' and charge specific quenching mechanisms should be investigated as the cause of the lack of optical activity in experimental characterizations. 

\end{abstract}
\maketitle

The nitrogen-vacancy centre in diamond is an attractive platform for quantum information processing~\cite{Wrachtrup_2006}, quantum sensing~\cite{Casola2019} and for storing quantum information~\cite{PhysRevX.9.031045}, and has been widely studied theoretically~\cite{Gali2019}.

The study of the NV centre in diamond has been dominated by the negatively and neutrally (\nvzero) charged states. This is due to the relative stability these charge states have in typical situations where the NV centre is found, for example in nanodiamonds in solution or substrate, or for diamond in bulk. The \nvminus\ is a promising candidate for singly addressable spin due to its excellent opto-mechanical properties and has demonstrated spin initialization and readout with a long spin coherence time.

Recently, the \nvplus\ charged state has attracted interest as a potential way to access long-lived nuclear spin states for spin coherence storage due to its absence of electronic spin~\cite{Pfender2017}. Furthermore, there is interest in studying alternate charge states of \nvminus\ as efficient readout of a single spin state, long term classical data storage, and high resolution microscopy~\cite{Han2010, Shields2015, Dhomkar2016}. 
However, these studies typically use \nvzero\ as it can be created with photoionization and easily identified optically.

Optically, \nvzero\ and \nvminus\ are very easy to see under typical experimental conditions via the optical absorption/emission spectrum due to a distinctive zero phonon line and phonon side band. In comparison, the \nvplus\ state is not thermodynamically favourable. Specifically, the \nvplus\ state is only stable for Fermi levels lower than approximately 1\,eV above the VBM, however N-doped diamond has a Fermi level of around 2\,eV above the valence band maximum (VBM)~\cite{Deak2014}. In order to fabricate stable \nvplus\ experimentally, it is necessary to reduce the Fermi level.

Experimental work towards creating the \nvplus\ state has successfully implemented both band bending, which increases the bands and relatively reduces the Fermi level, with functionalization of the diamond surface chemically~\cite{Hauf2011}, and directly controlling the Fermi level with in plane gates~\cite{Hauf2014,Pfender2017}, as well as a combination of both~\cite{Grotz2012,Schreyvogel2015}. These experimental papers reached a Fermi level that theoretically should stabilize \nvplus\ but only detected background optical emission. This is problematic, as the exact charge transition level varies by $\sim$0.5\,eV in theoretical estimates~\cite{Grotz2012,Weber2010,Deak2014,Londero2018}, so this technique cannot guarantee the existence of the \nvplus\ charge state. More concretely, Pfender et al.~\cite{Pfender2017} have measured an increased amplitude Rabi Oscillations of the nuclear spin, consistent with computational predictions that \nvplus\ should have no total electron spin~\cite{Meara2019}.

The lack of an optical signature makes the \nvplus\ difficult to identify and characterize. While hyperfine constants have been calculated~\cite{Meara2019} suggesting potential avenues of magnetic identification, no current optical identification is known and this state is often considered dark due to the lack of optical activity. 
This lack of optical activity is considered characteristic of the \nvplus\ state, however there exist optical transitions in the \nvplus\ that are dipole and spin allowed~\cite{Meara2019}. The experimentally observed dark state suggests there may be external quenching of the photoluminescence (PL) due to non-radiative transitions~\cite{Meara2019, Pfender2017}, and in~\cite{Schreyvogel2015} it is theorised that the PL is quenched by proximity to the two-dimensional hole accumulation layer created by functionalizing the surface that quenches \nvplus\ selectively. 

Calculating photoluminescence is generally limited by the computational complexity of quantum chemistry.
A solid state approach uses Delta self consistent field ($\Delta$-SCF) with the supercell method which has found success with the vibrational lineshape of the PL of a defect in bulk, but is only theoretically valid for excited states with different symmetry from the ground state~\cite{Gorling1999}. Cluster calculations allow us to use time-dependent density functional theory (TD-DFT) to obtain accurate ZPL and transition dipoles~\cite{Reimers2018,Gali2009b,gali2011b} but the result does not apply to solid state systems.
Recently, we have demonstrated a method based on cluster calculation with frequency cutoff that can accurately predict the ZPL and phonon sidebands of solid state systems~\cite{karim2020}.

In this work we report the first \textit{ab-initio} photoluminescence calculation showing that the \nvplus\ state is optically active and has a characteristic emission and absorption spectra. 

We perform TD-DFT calculation of the energy levels of the multielectron states (singlet and triplet states), and find that the transition dipole moments of the lowest singlet-singlet transition is non-zero.
Using our recently developed frequency cutoff method\cite{karim2020}, we predict the full PL spectrum. This suggest that this state cannot be considered ’dark’ but there might be another effect preventing experimental observation of the optical signature.

Our methodology is based on the frequency cutoff approach, recently reported in \cite{karim2020}. We use their nanodiamond shape, exchange-correlation functional and basis set. In brief, we calculate the photoluminescence (PL) spectra of a defect in bulk by applying a frequency cutoff to cluster calculations. The PL of the cluster is given under the displaced harmonic oscillator model. The only inputs for this model are the relaxed ground state, found with DFT; the relaxed excited state, found with TD-DFT; and the normal modes given by finite-difference under DFT of the ground state. This gives us the PL of the defect in a nanodiamond. To recover the solid state spectrum, this is followed by a TD-DFT calculation while constraining the outer layer of the cluster, which allows us to identify normal modes that occur on the surface, unique to the cluster. We eliminate this region of normal modes with a cutoff to recover the bulk PL spectrum.

To study the \nvplus\ center, we construct a 1\,nm diameter nanodiamond by adding to the vacancy the four nearest-neighbour carbon atoms in their bulk position and removing atoms from the outer layer to achieve C$_{3v}$ symmetry. We protonate the outer carbons to have consistent sp3 hybridisation and only CH and CH2 functional groups, which gives the chemical formula: C$_{197}$NH$_{140}$. 

\begin{figure}[t!]
    \centering
    \includegraphics[width=0.8\columnwidth]{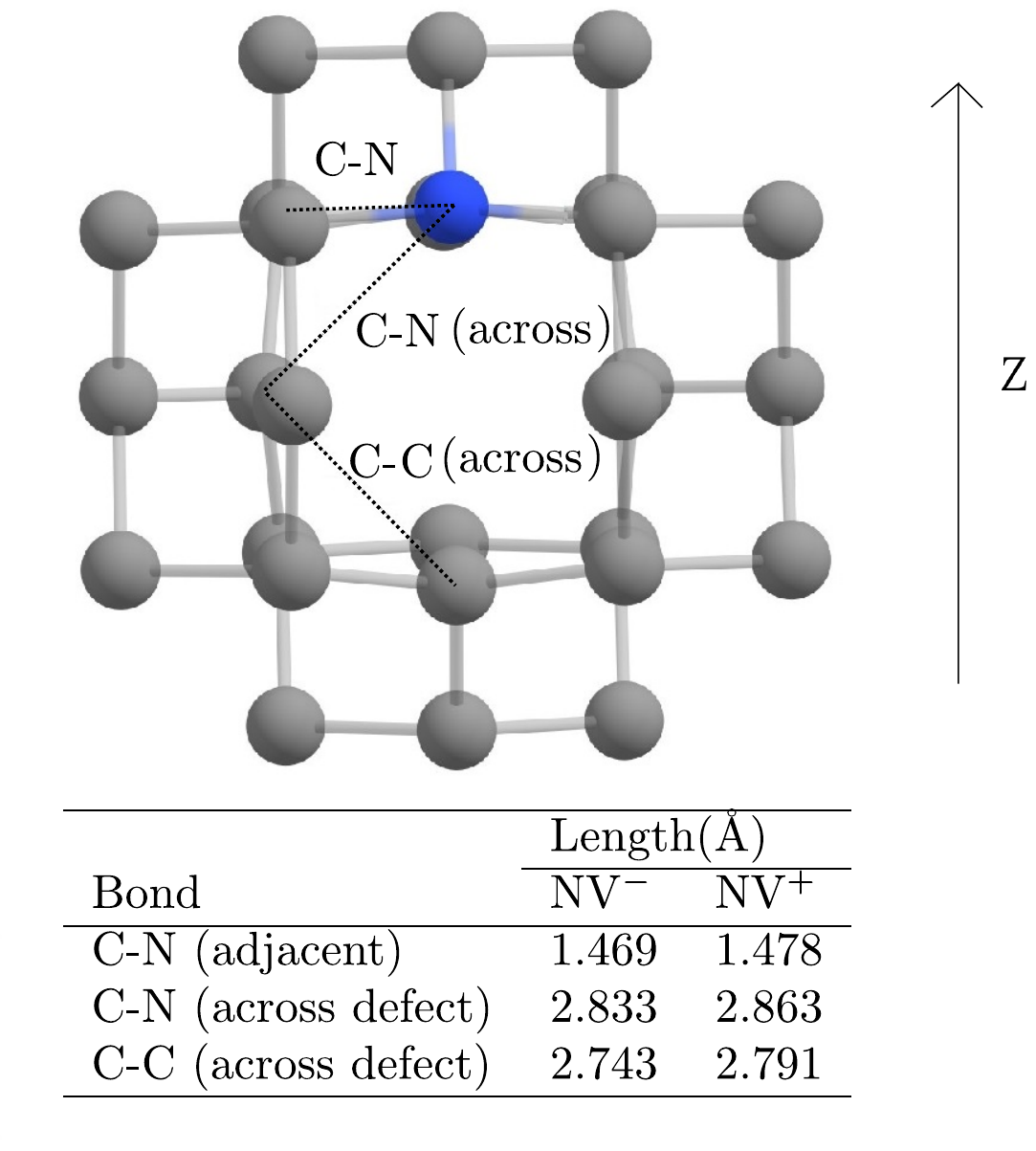}
    \caption{The ground state structure of \nvplus. Carbon atoms are shown in grey, and nitrogen atoms in blue (only the atoms around the defect are shown). The table reports the distance in \AA\ for atoms in the defect for \nvplus\ and \nvminus. Due to the reduced negative charge in the defect, the positive charged state has larger separations around the vacancy compared to the negatively charged state. Z axis for C$_{3v}$ geometry is shown and corresponds to the direction from the centre of the vacancy to the nitrogen atom.}
    \label{fig:structure}
\end{figure}

\begin{figure*}
    \centering
    \includegraphics[width=1\textwidth]{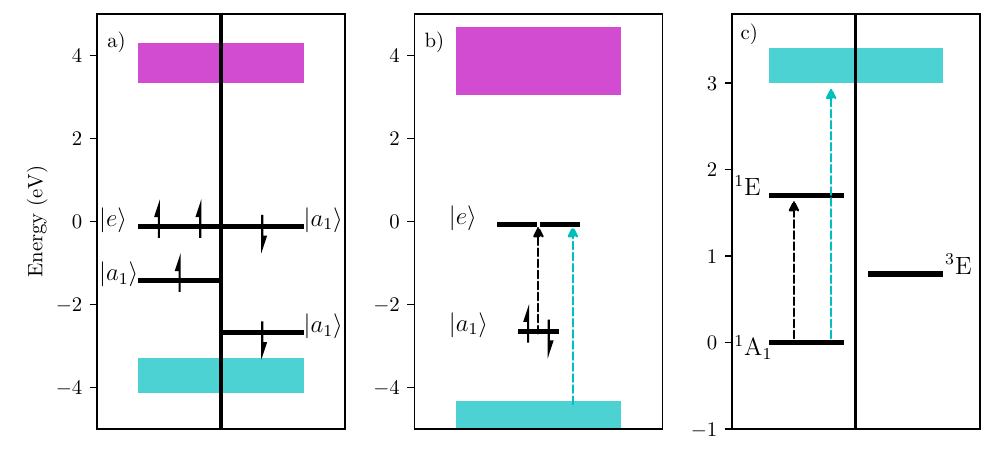}
    \caption{Kohn-Sham single electron orbital energies for \nvminus~(a) and \nvplus~(b). The multielectron energy ladder for \nvplus\ singlet (left) and triplet (right) are shown in c). Part a) shows alpha (left) and beta (right) electron orbitals separately since electrons are unpaired in \nvminus\.  Fishhooks represent electron occupation.  All multielectron states below 2 eV are intraband transitions; all higher states up to 4 eV are transitions from the valence band into the $\ket{e}$ state. Our proposed bright transition is shown as a black arrow in b) and c). The lowest singlet valence band transition occurs at 3.09 eV and is shown as a cyan arrow. It should be noted the virtual Kohn-Sham orbitals do not correspond to actual energies if they were occupied, so the bandgap cannot be read from a) or b) and excitation energies must be read from c). }
    \label{fig:electron}
\end{figure*}

Next, the ground state is relaxed from the bulk positions and the electron density and associated ground state energy under DFT in TURBOMOLE~\cite{Ahlrichs1989,Bauernschmitt1996} are converged for the stationary atoms. Throughout the process the symmetry is fixed to C$_{3v}$. Consistent with previous calculations with \nvminus\ \cite{karim2020}, we use def2-SV(P)~\cite{Schafer1992} as a basis set with PBE0~\cite{Perdew1996,Perdew1998} as the exchange-correlation functional. This allows explicit comparisons with \nvminus\ calculations.

Figure \ref{fig:structure} shows the ground state structure of \nvplus\ as well as the bond lengths and distances between the relevant atoms in the vacancy.
For \nvplus, the lack of the electron and consequent decreased negative charge within the vacancy compared to \nvminus\ increases the electrostatic repulsion between nuclei, which results in a larger space around the vacancy. The conformal change in comparison to \nvminus\ will lead to different coupling of vibrational modes, and a distinct vibronic spectrum.

The excited state is simulated under Linear Response Time-Dependent DFT using the TURBOMOLE software~\cite{Grimme2002,Furche2005} with the same basis set and functional. For the excited states, the symmetry is fixed to C$_{s}$. \nvplus\ was initially relaxed under C$_1$ symmetry and found to adopt C$_{s}$ symmetry. Furthermore C$_{s}$ symmetry was found to give accurate excited state properties for \nvminus\ in the harmonic approximation~\cite{Thiering2017}.

As the \nvplus\ is diamagnetic, all electrons are paired by spin.
This allows for DFT to be calculated with the closed shell Kohn-Sham equation~\cite{KohnSham1965} rather than unrestricted spin resolved calculations as with \nvminus. 

The single electron orbitals are shown in Figure \ref{fig:electron}(a). Similar to \nvminus\ shown in Figure \ref{fig:electron}(b), there are clear conduction and valence bands and importantly, there are also intraband levels local to the defect centre. These results are consistent with and extend previous calculations reported in the literature \cite{Meara2019}. The conduction bands shown are from virtual Kohn-Sham orbitals which may not accurately reflect the actual conduction band energies. The transition energies are calculated in the multielectron description.

We calculate the multielectron state energy ladder and report it in Figure \ref{fig:electron}(c), the ground state is a singlet state with A$_1$ symmetry. For the excited states within the bandgap, there are two singlets states with A' and A" symmetry as well as a triplet E state. The singlet states are degenerate on absorption (with E symmetry), but have different adiabatic transitions due to the reduction of symmetry from C$_{3v}$ to C$_{s}$ in the excited state geometry. All higher excited states involve exciting electrons  from the valence band to the empty e levels. The energy levels are reported in Figure~\ref{fig:transition}. As TD-DFT is used, the associated transition dipoles are calculated as reported in Figure~\ref{fig:transition}.

To access the singlet-triplet transition, it is necessary to use the Tamm-Dancoff Approximation (TDA) to avoid instabilities. The TDA ignores contribution by "de-excitation" terms i.e. the coupling from virtual to occupied orbitals, and has been shown to be valid for determining energies of single-excitation states, particularly for optical spectra~\cite{Chantzis2013, Peach2011}. Comparison of TDA transition energies for the singlet state are shown in supplementary material section I.

To obtain all the transition energies and strengths, we analyze the combined relaxed ground and excited states under the displaced harmonic oscillator model.  The results, reported in Figure \ref{fig:transition}, show that, in contrast to common conclusions reported in the literature, the transition dipole moments, $\mu$, are not zero and about a quarter of the value of \nvminus, which implies that the \nvplus\ is not a dark state. Furthermore, the zero phonon lines (ZPL) for the \nvplus\ states are lower in energy but under the phonon sideband of the \nvminus. This could explain why, to date, experimentally photon emission from this state has not been observed \cite{Hauf2011, Hauf2014, Grotz2012, Schreyvogel2015, Deak2014}.

\begin{figure}
    \centering
    \includegraphics[width=1\columnwidth]{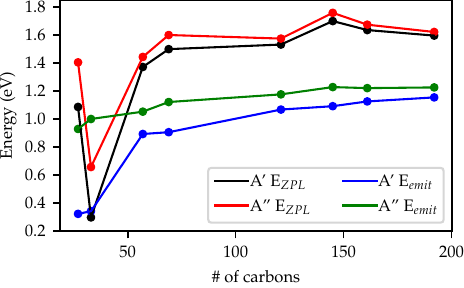}
    \caption{Emission energies from A' and A" states for different nanodiamonds. ZPL refers to the zero phonon line; Emi refers to the Vertical Emission. Only nanodiamonds with a majority of bond angles at $109.7^{\circ}$) are shown. Convergence to a specific value is not seen, however there is oscillation in a range of 0.1 eV, which we can use as error. Variation is due to shape dependence. The nanodiamonds have composition C$_{27}$NH$_{36}$, C$_{33}$NH$_{36}$, C$_{57}$NH$_{60}$, C$_{69}$NH$_{60}$, C$_{161}$NH$_{116}$}
    \label{fig:converge}
\end{figure}

\begin{figure}[t]
    \centering
    \includegraphics[width=\columnwidth]{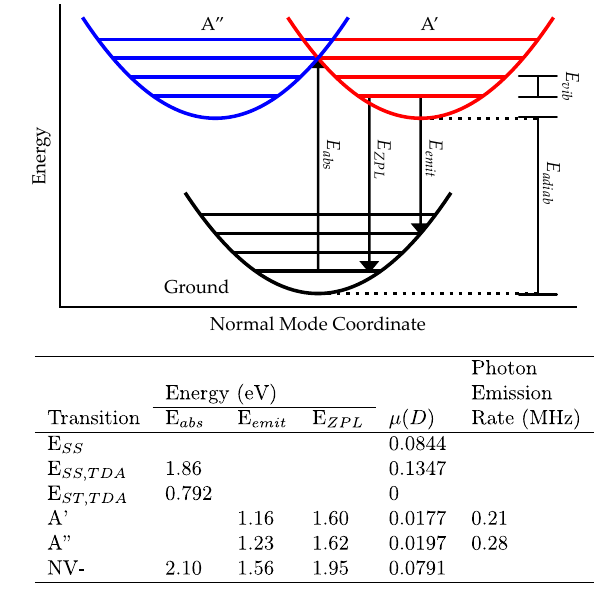}
    \caption{Transition energies between ground and excited states of \nvplus. $E_{SS}$ is the singlet-singlet transition, the Tamm-Duncoff Approximation (TDA) is used to calculate the singlet-triplet transition $E_{ST,TDA}$, and we also report the approximated singlet-singlet transition $E_{SS,TDA}$ for comparison. The ground state calculation gives the vertical absorption, $E_{abs}$.
    The relaxed excited state for A' and A" states give us the vertical emission, $E_{emit}$. The difference in energy between relaxed excited and relaxed ground states gives the adiabatic energy $E_{adiab}$. In the displaced harmonic oscillator model, $E_{vib}$ has the same value for ground and excited states, so $E_{adiab}$ has the same value as $E_{ZPL}$. The transition dipole, $\mu$ for each transition is given in Debye. The photon emission rate is calculated as the spontaneous emission rate from the Einstein A coefficient~\cite{Hilborn1998}.}
    \label{fig:transition}
\end{figure}

We consider convergence with nanodiamond size. Unfortunately, all energetics are highly dependent on the shape of the nanodiamond and, furthermore, larger nanodiamond sizes can form graphite on the surface~\cite{Barnard2003}. Figure \ref{fig:converge} shows emission values for select sizes where the cluster retained diamond geometry (majority of bond angles are $109.7^{\circ}$). The ZPL can be seen to flatten after 121 carbon atoms and oscillate by around 0.1 eV around their central values.

\begin{figure*}
    
    \vspace*{-3mm}
    \subfloat{\includegraphics{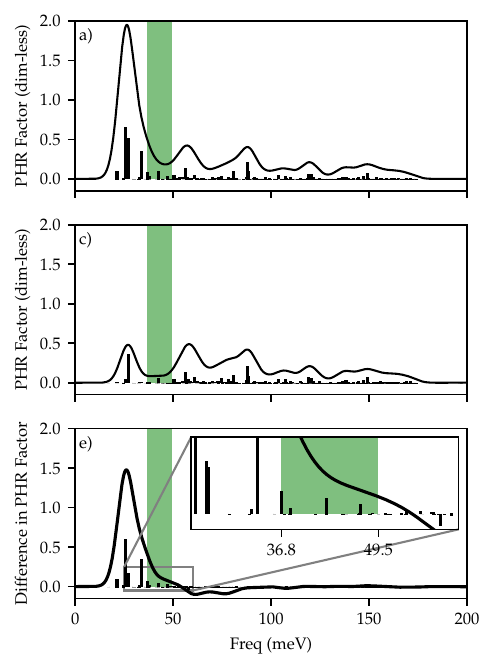}}
    \subfloat{\includegraphics{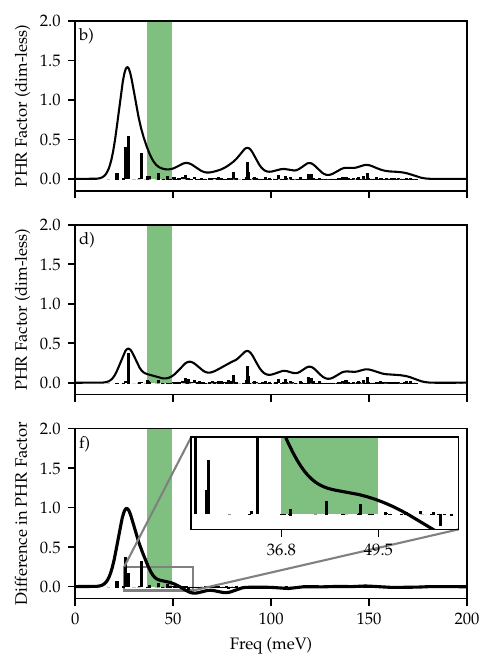}}
    
    \caption{Partial Huang-Rhys (PHR) factors as functions of frequency for unconstrained (a,b) and constrained (c,d) C$_{197}$NH$_{140}$ nanodiamond. A' excited state results are given in the left column as indicated by a). Similarly, A" is given on the right. The black vertical bars are the PHR factors and the solid black line shows gaussian broadening to identify regions of peaks. The cutoff region~\cite{karim2020} is shaded green from 36.8\,meV to 49.5\,meV. In the constrained calculation, frequencies below this region are suppressed while those above this region are preserved. The cutoff frequency value is selected within this region to calculate the PL spectrum.}
    \label{fig:HR-NVP}
\end{figure*}

We now analyse the vibrational properties of the \nvplus\ defect. Vibrational modes for the ground state are calculated with finite-difference under DFT with SNF software \cite{Neugebauer2002}. We can then calculate the Partial Huang-Rhys (PHR) factors for each vibrational mode and each transition, which show how each vibrational mode couples with a transition. The results are shown in Figure~\ref{fig:HR-NVP}(a) for A' and (b) for A" respectively. 

The vibrational modes unique to the nanodiamond, can be found by applying the same constraints present in the bulk by calculating the relaxed excited state while constraining the outer layer of carbons in the nanodiamond, as detailed in~\cite{karim2020}. This new excited state and the ground state vibrational modes are used in VIBES to compute the constrained PHR factors shown in Figure~\ref{fig:HR-NVP}(c) and (d) for A' and A" transitions respectively.

The difference between unconstrained and constrained spectra is plotted in Figure~\ref{fig:HR-NVP}(e) and (f). Here we can see that at low frequencies (on the left of the green shaded area) the presence of large positive peaks indicate that the coupling to vibrational modes are present in the nanodiamond but not in bulk, whereas at high frequencies (on the right of the green shaded area) the vibrational modes equally couple to the transition for both bulk and nanodiamond. The green shaded area is therefore identified as a region separating the unique properties of the nanodiamond from those in common with bulk. Eliminating the vibrational modes with frequencies below this  cutoff region should return the bulk behaviour.
The lower boundary of the cutoff region is at 36.8\,meV, and the upper boundary is at 49.5\,meV. Choosing a cutoff value within this region has been shown to reproduce a simulated PL consistent with experimental measurements \cite{karim2020}. 

\begin{figure*}[t]
    \centering
    \includegraphics[width=1\textwidth]{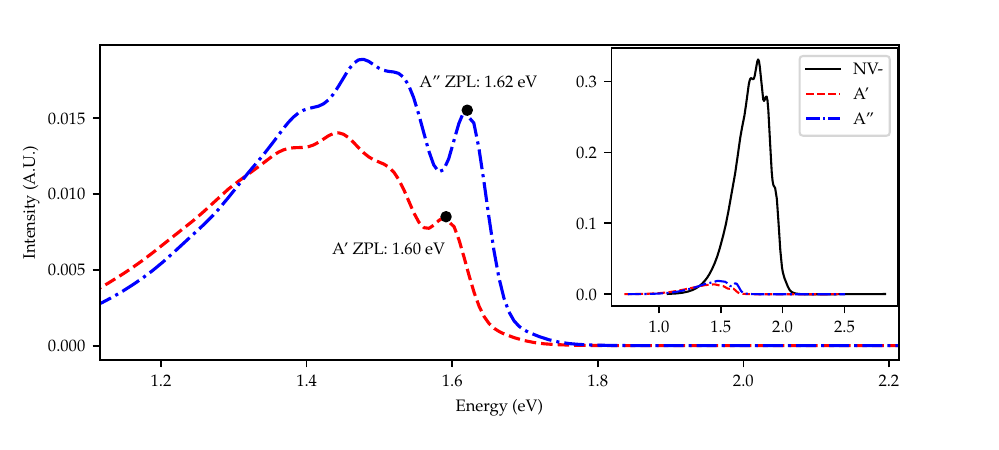}
    \caption{Predicted photoluminescence of the two excited states of the \nvplus\ defect in diamond calculated in the middle of the cutoff regions. The intensity of emission should scale as the square of the transition moment. The inset compares the intensity of \nvplus\ with the \nvminus\ emission~\cite{Aslam2013} which is $\sim$16x higher than that of \nvplus. Experimental data~\cite{Aslam2013} \copyright IOP Publishing and Deutsche Physikalische Gesellschaft  Reproduced by permission of IOP Publishing. CC BY-NC-SA}
    \label{fig:PL}
\end{figure*}

The large peaks in the PHR factors of \nvminus\ have been shown to be directly responsible for the main features of the PL spectrum. 
We compare the PHR factors for \nvplus\ reported in Figure \ref{fig:HR-NVP} to those reported for \nvminus\ in~\cite{karim2020}. \nvminus\ exhibits a large peak in the low frequency region, at 30\,meV which is reduced under constrained calculation. Similarly the \nvplus\ has a large peak at 25\,meV. In the higher frequencies, \nvminus\ has a large peak unaffected by constraining, centered at 66~meV, whereas \nvplus\ has two peaks, one at 60\,meV and one at 80\,meV associated with the A' and A" transitions respectively. It can also be seen that there is non-negligible coupling at frequencies up to 175\,meV, compared to \nvminus\ for which it becomes negligible above 100\,meV.

To generate the PL spectra, we apply the cutoff at the center of the region (43.2\,meV) to the unconstrained nanodiamond calculations. The modified PHR values are used in VIBES to get a vibrational lineshape, which we center on the ZPL from Figure~\ref{fig:transition} to get the PL.

Figure \ref{fig:PL} shows the PL spectra for the two transitions A' and A". 
The intensity of A" is higher than A' as expected from the larger transition dipole moment as shown in Figure \ref{fig:transition}, but both lower than \nvminus by a factor of around twenty. The photon emission rate is calculated as the Einstein A coefficient~\cite{Hilborn1998}. The ZPL is at $1.60 \pm 0.1$ and $1.62 \pm 0.1$\,eV for A' and A" respectively, which are both lower energy than $1.95$\,eV of \nvminus\ and within the phonon sideband of \nvminus.
Experimentally, the emission spectrum should match the A' curve (dashed red line) as it is more energetically favourable for the excited state to deform to the A' configuration (longer wavelength).

It has been proposed in previous literature that the defect charge transition from \nvzero~to \nvplus\ at 1 eV above the VBM indicates photoionisation may occur instead of bright transitions of higher energy~\cite{Meara2019}. The defect charge transition state is calculated by assuming the defect is coupled to an electron bath with some chemical potential. The transition state is defined as the chemical potential of the bath where the defect formation energy for two charge states are identical. This corresponds to the energy difference between the two charge states when the chemical potential of the bath is at the CBM. The defect transition energies can be considered the chemical potential of the bath required for 50\% population of each charge state. This model adequately explains thermodynamics of charge states. Experimentally, when the Fermi level (the chemical potential of the bath) is tuned to cross the defect transition energies, the population of one charge state changes to the next as indicated by change in photoluminescence~\cite{Pfender2017}. 

However, investigation into \nvplus/\nvzero\  extrinsic optical dynamics shows radiative transitions that cannot be explained under this model~\cite{Doherty2013}. There exist both radiative and non-radiative transitions between charge states. Non-radiative transitions consist of electrons tunnelling between defect centres and electron donors/acceptors and can be controlled by isolating the defect. Radiative transitions involve electrons entering or exiting the defect via the bulk bands and require explicit calculation of defect to band transitions. Explicit calculation of \nvzero/\nvminus\ radiative photoionisation transition energies has recently been performed by Razinkovas~\cite{Razinkovas2021} using defect to band transitions. They calculate the energy difference between \nvminus\ and \nvzero\ with one electron excited to the conduction band. They found the energy difference of the ground state of \nvminus\ to \nvzero\ with one CBM electron gave the experimental threshold for photoionisation of 2.64 eV.

Our equivalent transition is between the ground state of \nvplus\ with \nvzero\ with one fewer electron in the valence band. Under TD-DFT we can simulate this by relaxing the excited state with one electron excited from the valence band, which is shown by the cyan arrow in Figure~\ref{fig:electron}(c). We perform this calculation in C$_{s}$ geometry, Similar to the transition to the $^{1}$E state, however now we optimise for the second excited A' and A" states. The energies are 3.09 and 3.14 eV respectively, which are higher than our proposed bright ZPL of 1.7 eV. Illumination by 1.7 eV light should be able to drive the bright transition within the defect without inducing this charge conversion transition.

\def\arraystretch{1.5}
\begin{table}[]
\caption{Energy values as a function of the applied electric field in the z direction. Gap refers to the Homo-Lumo Gap. Value is the absolute value. Diff is the difference between the energy of the \nvplus\ under applied field and the energy under the zero electric field.}
% \vspace{12 pt}
\begin{tabular}{l|lll|l}
\hline
\multirow{2}{*}{\begin{tabular}[c]{@{}l@{}}Electric Field\\ {[}kV/cm{]}\end{tabular}} & \multicolumn{2}{l|}{Gap {[}eV{]}}                          & \multicolumn{2}{l}{Ground Energy {[}eV{]}} \\ \cline{2-5} 
                                                                                      & \multicolumn{1}{l|}{Value}  & \multicolumn{1}{l|}{Diff}   & Value                   & Diff              \\ \hline
-100                                                                                  & \multicolumn{1}{l|}{2.4645} & \multicolumn{1}{l|}{0.0071} & -207661.96706           & -0.00348          \\
100                                                                                   & \multicolumn{1}{l|}{2.4787} & \multicolumn{1}{l|}{0.0071} & -207661.97521           & 0.00467           \\
0                                                                                     & \multicolumn{1}{l|}{2.4716} & \multicolumn{1}{c|}{-}      & -207661.97054           & 0                 \\ \hline
\end{tabular}\label{table:elec}
\end{table}

There is also a transition from the excited $^{1}$E state to a possible \nvzero state, by first exciting from $\ket{a_1}$ to $\ket{e}$, and then from the valence band to $\ket{a_1}$ as shown in Figure~\ref{fig:electron}(b). This transition has an energy of 1.3 eV and may result in charge state conversion. However, since it requires two excitations, it will scale second order with power, and the timescales will be much longer than transitions within a charge state as is the case for the equivalent second order \nvminus/\nvzero\  transition~\cite{Doherty2013}. Therefore, it should be theoretically possible to see PL without resulting in immediate photoionisation. For further details see supplementary information section II.

As \nvplus\ measurements often occur under electric field~\cite{Pfender2017, Hauf2014, Grotz2012,Schreyvogel2015}, we consider the effect of the electric field on the energy of the defect. Hauf et al.~\cite{Hauf2014} experimentally observed \nvplus\ under an electric field strength of 100\,kV/cm. 
To understand the effect of the electric field on the ground state energy, we apply a linear electric field and perform a relaxation under DFT in TURBOMOLE, to get the new ground state energy and associated deformation for those values. 
The calculation results are reported in Table~\ref{table:elec}, where we see that applying an electric field of 100\,kV/cm shifts the homo-lumo gap by only 7\,meV, which suggests that the expected effect on the PL spectra and photoionisation is small.

We have demonstrated that the \nvplus\ state contains an optically active transition between its singlet ground and singlet excited levels. 
We have studied the emission both electronically and vibrationally, calculating the TD-DFT energy levels for the excited triplet state and found that the dipole transition moment is non-zero. Finally, we have predicted the characteristic photoluminescence spectra and shown that the effects of the electric field are negligible. These results are consistent across three sizes of nanodiamond. While our results appear to be in contradiction with recent experimental observations of \nvplus\ concluding that it is a dark state, this could be accounted for by a charge dependent quenching mechanism.

\vspace{5mm}
\noindent\textbf{DATA AVAILABILITY} \\
The data that support the findings of this study are available from the corresponding author upon reasonable request.

\vspace{5mm}
\noindent\textbf{ACKNOWLEDGMENTS} \\
We thank Cristian Bonato, Alastair Stacey and Brett Johnson for useful discussions. 
A.P. acknowledges a RMIT University Vice-Chancellor’s Senior Research Fellowship; ARC DECRA Fellowship (No: DE140101700), and Google Faculty Research Award. This work was supported by the Australian Government through the Australian Research Council under the Centre of Excellence scheme (No: CE170100012, CE170100026). It was also supported by computational resources provided by the Australian Government through the National Computational Infrastructure Facility and the Pawsey Supercomputer Centre.

\setcounter{table}{0}
\renewcommand{\thetable}{S\arabic{table}}%
\setcounter{figure}{0}
\renewcommand{\thefigure}{S\arabic{figure}}%
\renewcommand{\theequation}{S\arabic{equation}}

\vspace{5mm}
\noindent\textbf{SUPPLEMENTARY MATERIAL I: Tamm-Dancoff Approximation} \\
The entire methodology was performed for \nvplus\ in nanodiamonds of three different sizes: C$_{121}$NH$_{100}$, C$_{145}$NH$_{100}$ and C$_{197}$NH$_{140}$. In the main text we showed only results for the largest nanodiamond, and in this section we report values for the other sizes.
Table~S1 shows the transition energies and strengths for all three sizes of nanodiamond. 
It can be observed that the Tamm-Dancoff Approximation (TDA) overestimates the singlet transition energies by 10\% for all sizes. The dipole moments of C$_{121}$NH$_{100}$ and C$_{145}$NH$_{100}$ diamond are overestimated by an order of magnitude.
For singlet states, we use TD-DFT without TDA, however for the triplet states we use TDA due to triplet instabilities. The singlet-triplet transition energy is $0.79\pm0.08$ eV, where we estimate the error, from the singlet-singlet result, to be about 10\%. Singlet-triplet transitions are dipole forbidden, so there is no need to calculate the transition dipole moments for the triplet state.

There is a second triplet state transition at $2.80$ eV that corresponds to exciting from the valence band into the defect in a triplet E state. There is no radiative transition from the ground state into this state. The transition between triplet ground state and this valence band state is approximately $2$ eV.

\begin{table}[b]
\caption{Energy comparison of nanodiamonds of sizes C$_{121}$NH$_{100}$, C$_{145}$NH$_{100}$ and C$_{197}$NH$_{140}$. Transitions are labelled with symmetry label and then subscript SS for singlet -singlet; ST for singlet-triplet; and ST2 for the second highest singlet-triplet state. This involves exciting from the valence band into the defect.}
\begin{tabular}{lllll}
\hline
&\multicolumn{2}{l}{Energy (eV)} & & \\\cline{2-4}
Transition & E$_{abs}$ & E$_{emit}$ & E$_{ZPL}$  & $\mu$ (D)\\
\hline
E$_{SS,197CN}$      & 1.70       &    -        &    -        & 0.0844  \\ 
E$_{SS,TDA,197CN}$  & 1.86       &      -      &       -     & 0.1347  \\ 
E$_{ST,TDA,197CN}$  & 0.79      &    -        &       -     & 0       \\
E$_{ST2,TDA,197CN}$  & 2.80      &    -        &       -     & 0       \\
A'$_{197CN}$          &     -      & 1.16       & 1.60        & 0.0177   \\ 
A"$_{197CN}$          &     -      & 1.23       & 1.62        & 0.0197  \\
\hline
E$_{SS,121CN}$      & 1.72       &    -        &    -        & 0.0519   \\
E$_{SS,TDA,121CN}$  & 1.87       &    -        &        -    & 0.90741   \\
A'$_{121CN}$          &     -      & 1.06       & 1.62        & 0.0101  \\
A"$_{121CN}$          &     -      & 1.17       & 1.64        & 0.0119   \\
\hline
E$_{SS,145CN}$      & 1.83       &    -        &    -        & 0.0496  \\
E$_{SS,TDA,145CN}$  & 1.96       &      -      &       -     & 0.8128  \\ 
A'$_{145CN}$          &     -      & 1.10        & 1.70        & 0.0092 \\
A"$_{145CN}$          &     -      & 1.23        & 1.76        & 0.012  \\
\hline
\nvminus         &     2.10   &    1.56     &    1.95     & 0.0791  \\ \hline 
\end{tabular}
\end{table}\label{table:othersize}

\begin{figure*}
    \centering
    \includegraphics[width=\textwidth]{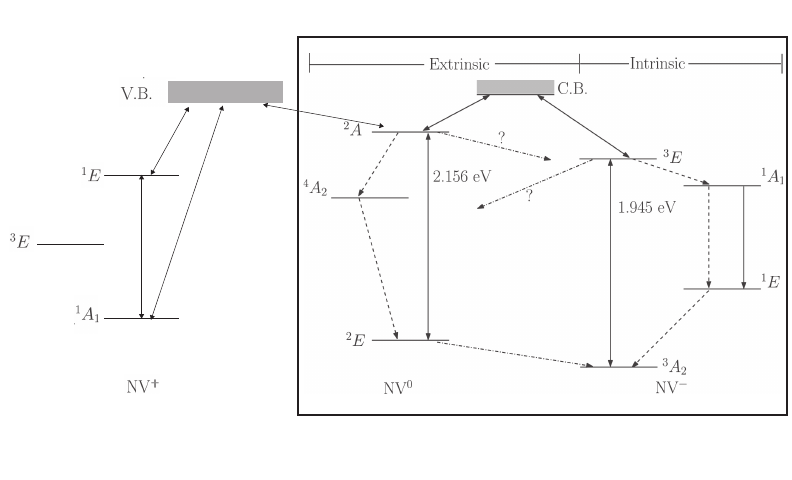}
    \caption{ Diagram of transitions between the charge states of the NV centre in diamond. The transitions between the \nvzero\ and \nvminus, shown inside the black border, are reproduced from Doherty et al.~\cite{Doherty2013} with permission from Elsevier. The radiative charge state transition from \nvminus\ to \nvzero\ occurs first through excitation to the conduction band. In addition, we propose that for \nvplus, assuming no electrons in the conduction band, the only available radiative mechanism occurs through an excitation from the valence band (``V.B.'').}
    \label{fig:extrinsic}
\end{figure*}

\vspace{5mm}
\noindent\textbf{SUPPLEMENTARY MATERIAL II: Charge State Conversion} \\
Figure~\ref{fig:extrinsic} shows the multielectron energy levels and subsequent charge state transition between \nvminus, \nvzero, and \nvplus. There are intrinsic dynamics within a charge state and extrinsic dynamics that allow transitions between charge states~\cite{Doherty2013}. Non-radiative transitions involve tunnelling of an electron between the NV centre and a nearby well, typically an N defect or a functional group on the surface and vary according to proximity to the defect~\cite{Doherty2013}. These are shown as dotted arrows in Figure~\ref{fig:extrinsic}. Radiative transitions, however, involve excitations of the bulk diamond band structure and are due to the defect itself. Assuming isolation of the defect, under illumination, charge transitions will occur primarily through radiative transitions.

A radiative excitation from \nvminus\ to \nvzero\ occurs through the conduction band. An electron from \nvminus\ can be excited into the conduction band leaving the defect in the \nvzero\ charge state. The electron in the conduction band can then relax into another well such as a defect or surface functional group~\cite{Doherty2013}. The energy to excite an electron from the ground state is $2.6$ eV and can be seen experimentally~\cite{Doherty2013}. These are shown as solid arrows in Figure~\ref{fig:extrinsic}.

We propose a similar mechanism but involving the valence band for the transitions from \nvplus\ to \nvzero. In our case, we assume an isolated \nvplus\ defect with no electrons in the conduction band under illumination. The optical transitions available are given as solid arrows in Figure~\ref{fig:extrinsic}. We propose that under illumination, an electron can be promoted from the valence band into the defect. The hole in the valence band can then be filled elsewhere in the crystal similar to how the conduction band electron is assumed to relax elsewhere in the \nvminus\ to \nvzero\  transition~\cite{Doherty2013}.

Our proposed intrinsic dynamics involve a bright transition between the ground $^1$A$_1$ state and $^1$E singlet states at $1.7$ eV. We also see a triplet $^3$E state in-between the singlet states. 

We also propose bright transitions where a valence band electron is promoted to a defect level. These are shown as a grey box labelled ``V.B.'' in Figure~\ref{fig:extrinsic}. For more details, see main text Figure 2(b) and 2(c). The transition from the ground state to the lowest band state is $3.08$ eV. We propose this corresponds to the equivalent \nvminus\ transition at $2.6$ eV. 

The second order transition requires exciting the defect to the singlet E state before exciting into the valence band. The second excitation should take $1.38$ eV. It is possible for this transition to occur under illumination of $1.7$ eV light, however, since it is second order, it should occur at a longer timescale than the intrinsic dynamics~\cite{Doherty2013} and, more importantly, scale as the square of input power. Therefore, this effect will be suppressed at low power. Calculating the transition dipole moment between two excited states is outside of the scope of this paper, so the actual rates cannot currently be compared. 

If the defect is in its triplet ground state, there is also an optical transition of around $2.0$ eV. This, similarly, will occur at long timescales since \nvplus\ must change spin states.

\vspace{5mm}
\noindent\textbf{REFERENCES}
\bibliography{NVPL}

\end{document}